\documentclass[11pt]{amsart}
\usepackage{amsmath}
\usepackage{amsfonts}
\usepackage{amssymb}
\newtheorem{thm}{Theorem}

\newtheorem{remark}[thm]{Remark}
\newtheorem{lemma}[thm]{Lemma}

\newtheorem{exam}[thm]{Example}
\newtheorem{defn}[thm]{Definition}

\newcommand{\ket}[1]{| #1 \rangle}

\newcommand{\ketbra}[2]{| #1 \rangle\langle #2 |}
\newcommand{\Tr}{{\rm Tr}}

\newcommand{\spn}{\operatorname{span}}
\newcommand{\fA}{\mathfrak{A}}
\newcommand{\fB}{\mathfrak{B}}

\begin{document}

\title[The Multiplicative Domain in Quantum Error Correction]{The Multiplicative Domain in \\ Quantum Error Correction}

\author[M.-D.~Choi, N.~Johnston, D.~W.~Kribs]{Man-Duen Choi$^1$, Nathaniel Johnston$^2$, and David~W.~Kribs$^{2,3}$}
\address{$^1$Department of Mathematics, University of Toronto, ON Canada M5S 3G3}
\address{$^2$Department of Mathematics \& Statistics, University of Guelph,
Guelph, ON, Canada N1G 2W1}
\address{$^3$Institute for Quantum Computing, University of Waterloo, Waterloo, ON, Canada
N2L 3G1}

\begin{abstract}
We show that the multiplicative domain of a completely positive
map yields a new class of quantum error correcting codes. In the
case of a unital quantum channel, these are precisely the codes
that do not require a measurement as part of the recovery process,
the so-called unitarily correctable codes. Whereas in the
arbitrary, not necessarily unital case they form a proper subset
of unitarily correctable codes that can be computed from
properties of the channel. As part of the analysis we derive a
representation theoretic characterization of subsystem codes. We
also present a number of illustrative examples.
\end{abstract}

\maketitle

\section{Introduction \& Preliminaries}

Quantum error correction lies at the heart of many investigations
in quantum information science \cite{NC00,Got02,KLABVZ02}. As
theoretical and experimental efforts become more ramified, and in
particular as attempts are made to bring the two perspectives
closer together, the need grows for techniques that can identify
error correcting codes for wider classes of noise models. Indeed,
whereas many approaches to quantum error correction rely on
special features of the noise operators under consideration, such
as the stabilizer formalism \cite{Got96} and group theoretic
properties of Pauli operators for instance, in the general setting
of Hamiltonian driven noise descriptions an arbitrary noise model
will in general have no tractable algebraic properties. Recent
work in quantum error correction has thus included considerable
effort toward the goal of identifying quantum codes for ever wider
classes of noise models. See
\cite{KLP05,Pou05,SL05,Bac05,KS06,CK06,Kni06,KlaSar06,AlyKla07,ESMRLBCL07,BKK07a,BNPV08,SMKE08}
and the references therein for a variety of discussions and
analysis.

In this paper we contribute to this line of investigation by
showing the multiplicative domain of a completely positive map, a
notion first studied in operator theory over thirty years ago
\cite{Cho74,Paulsentext}, yields a new class of quantum error
correcting subspace and subsystem codes. The multiplicative domain
codes form a subclass of what are known as ``unitarily correctable
codes'' \cite{KS06,SMKE08,KLPL06} (UCC). These are codes that do
not require a measurement as part of the recovery process, in
other words they are highly degenerate codes for which a unitary
recovery operation can be obtained. The UCC class also includes
decoherence-free subspaces and noiseless subsystems
\cite{zanardi97, palma96, duan97, lidar98,knill00, zanardi01a,
kempe01, CK06,Kni06}, and other special codes such as unitarily
noiseless subsystems \cite{BNPV08}.
Additionally, our analysis includes a derivation of a
representation theoretic description of subspace and subsystem
codes that we believe is of independent interest. Specifically, we
show every code can be characterized in the Schr\"{o}dinger
picture for quantum dynamics as a ``smeared'' representation. This
complements other recently obtained descriptions of subsystem
codes \cite{KS06,Kni06,BKP08,BKK08}.

\strut

Before moving to the core of the paper we briefly present our
notation and nomenclature.

For our purposes, $\mathcal{H}$ will be a finite-dimensional
Hilbert space, $\mathcal{L}(\mathcal{H})$ is the set of linear
operators on $\mathcal{H}$, and $\mathcal{L}_1(\mathcal{H})$
denotes the set of trace class operators. The latter two sets of
operators are isomorphic in the finite-dimensional case, and so we
will use this identification when convenient. In the
Schr\"{o}dinger picture for quantum dynamics, time evolution of
open quantum systems is described by completely positive (CP)
trace preserving maps $\mathcal{E}:\mathcal{L}_1(\mathcal{H})
\rightarrow \mathcal{L}_1(\mathcal{H})$; for which a family of
operators $\mathcal{E}\equiv\{E_i\}$ can be found with
$\mathcal{E}(\rho) = \sum_i E_i \rho E_i^\dagger$ for all
$\rho\in\mathcal{L}_1(\mathcal{H})$ and $\sum_i E_i^\dagger E_i =
I$. (Here we use $E^\dagger$ for the operator adjoint, or
conjugate transpose for matrices.) We refer to such a map as a
\emph{quantum operation} or \emph{channel}. On the other hand,
evolution in the Heisenberg picture is described by the dual map
$\mathcal{E}^\dagger : \mathcal{L}(\mathcal{H}) \rightarrow
\mathcal{L}(\mathcal{H})$ defined via $\Tr(\mathcal{E}(\rho)X) =
\Tr (\rho \mathcal{E}^\dagger(X))$. Observe that
$\mathcal{E}\equiv\{E_i\}$ if and only if
$\mathcal{E}^\dagger\equiv\{E_i^\dagger\}$, and $\mathcal{E}$ is
trace preserving if and only if $\mathcal{E}^\dagger$ is unital
($\mathcal{E}^\dagger(I)=I$).

Standard quantum error correction considers quantum codes as
subspaces $\mathcal{C}\subseteq\mathcal{H}$
\cite{Got96,shor95,steane96,bennett96}. The code $\mathcal{C}$ is
said to be \emph{correctable} for $\mathcal{E}$ if there is a
channel $\mathcal{R} :
\mathcal{L}_1(\mathcal{H})\rightarrow\mathcal{L}_1(\mathcal{H})$
such that $\mathcal{R}\circ \mathcal{E} \circ
\mathcal{P}_\mathcal{C} = \mathcal{P}_\mathcal{C}$, where
$\mathcal{P}_\mathcal{C}(\rho)=P_\mathcal{C} \rho P_\mathcal{C}$
and $P_\mathcal{C}$ is the orthogonal projection of $\mathcal{H}$
onto $\mathcal{C}$. Given $\mathcal{E}\equiv\{E_i\}$, the
Knill-Laflamme Theorem \cite{KL97a} shows $\mathcal{C}$ is
correctable for $\mathcal{E}$ if and only if there is a complex
matrix $\Lambda = (\lambda_{ij})$ such that $P_\mathcal{C}
E_i^\dagger E_j P_\mathcal{C} = \lambda_{ij} P_\mathcal{C}$ for
all $i,j$. Observe the matrix $\Lambda$ is necessarily a density
matrix; i.e., positive with trace equal to one.

A generalization called ``operator quantum error correction''
\cite{KLP05,KLPL06} leads to the notion of \emph{subsystem codes}
\cite{Pou05,Bac05,KlaSar06,AlyKla07}. Two Hilbert spaces
$\mathcal{A}$, $\mathcal{B}$ are \emph{subsystems} of
$\mathcal{H}$ when $\mathcal{H}$ decomposes as $\mathcal{H} =
\mathcal{C} \oplus \mathcal{C}^\perp$ with $\mathcal{C} =
\mathcal{A} \otimes \mathcal{B}$. Notationally, we shall write
$\rho_\mathcal{A}$ for operators in $\mathcal{L}_1(\mathcal{A})$,
etc. A subsystem $\mathcal{B}$ is \emph{correctable} for
$\mathcal{E}$ if there is a channel $\mathcal{R} :
\mathcal{L}_1(\mathcal{H})\rightarrow\mathcal{L}_1(\mathcal{H})$
and a channel $\mathcal{F}_\mathcal{A} :
\mathcal{L}_1(\mathcal{A})\rightarrow\mathcal{L}_1(\mathcal{A})$
such that $\mathcal{R}\circ \mathcal{E} \circ
\mathcal{P}_\mathcal{C} = (\mathcal{F}_\mathcal{A} \otimes {\rm
id}_\mathcal{B})\circ \mathcal{P}_\mathcal{C}$. An extension of
the Knill-Laflamme Theorem to subsystems \cite{KLP05,KLPL06,NP05}
shows $\mathcal{B}$ is correctable for $\mathcal{E}$ if and only
if there are operators $F_{ij}\in\mathcal{L}(\mathcal{A})$ such
that $P_\mathcal{C} E_i^\dagger E_j P_\mathcal{C} = (F_{ij}
\otimes I_\mathcal{B}) P_\mathcal{C}$, where $I_\mathcal{B}$ is
the identity operator on $\mathcal{B}$. This is equivalent to the
existence of a channel $\mathcal{F}_\mathcal{A}$ such that
$\mathcal{P}_\mathcal{C}\circ \mathcal{E}^\dagger \circ
\mathcal{E} \circ \mathcal{P}_\mathcal{C} =
(\mathcal{F}_\mathcal{A} \otimes {\rm id}_\mathcal{B})\circ
\mathcal{P}_\mathcal{C}$. As a notational convenience, given
operators $X\in\mathcal{L}(\mathcal{A})$ and
$Y\in\mathcal{L}(\mathcal{B})$, we will write $X\otimes Y$ for the
operator on $\mathcal{H}$ given by $(X\otimes Y) \oplus
0_{\mathcal{C}^\perp}$.

It is often convenient in quantum information to work in an
operator algebraic setting. For our purposes, an \emph{operator
algebra} $\fA$ will refer to a finite-dimensional von Neumann
algebra \cite{Dav96}; that is, a set of operators inside
$\mathcal{L}(\mathcal{H})$ that is closed under taking linear
combinations, multiplication, and adjoints. Every algebra $\fA
\subseteq \mathcal{L}(\mathcal{H})$ induces an orthogonal direct
sum decomposition of the Hilbert space $\mathcal{H} = \oplus_k
(\mathcal{A}_k \otimes \mathcal{B}_k) \oplus \mathcal{K}$ such
that the algebra $\fA$ consists of all operators belonging to the
set
\begin{equation}\label{opalgform}
\fA = \oplus_k \big(I_{\mathcal{A}_k} \otimes
\mathcal{L}(\mathcal{B}_k)\big) \, \oplus \, 0_\mathcal{K},
\end{equation}
where $0_\mathcal{K}$ is the zero operator on $\mathcal{K}$.

\section{Representation Theoretic Description of Subsystem Codes}

Suppose $\fA$ is
an operator algebra on a Hilbert space $\mathcal{H}$. By a
\emph{representation} or a \emph{$\ast$-homomorphism} of $\fA$, we
mean a linear map $\pi: \fA \rightarrow \mathcal{L}(\mathcal{H})$
that is multiplicative and preserves the adjoint operation:
\begin{eqnarray*}
\pi(ab)&=& \pi(a)\pi(b) \quad\quad \forall a,b\in\fA \\
\pi(a^\dagger) &=& \pi(a)^\dagger \quad\quad\quad  \forall a\in\fA
\end{eqnarray*}
Every representation $\pi$ of $\fA = 1_n\otimes
\mathcal{L}(\mathcal{H})$, where $\mathcal{H}$ is
finite-dimensional, has a very special form \cite{Dav96}: there is
a positive integer $m$ and a unitary $U$ from
$\mathcal{H}^{\otimes m}$ into the range Hilbert space for $\pi$
such that
\begin{eqnarray}\label{repnform}
\pi(a) = U (1_m\otimes a) U^\dagger \quad\quad \forall a\in\fA.
\end{eqnarray}
The integer $m$ is referred to as the \emph{multiplicity} of the
representation $\pi$. In what follows, we will apply this
representation theory to the algebras $\mathcal{L}_1(\mathcal{C})$
and $\fA_\mathcal{B}:=
1_\mathcal{A}\otimes\mathcal{L}_1(\mathcal{B})$.

\subsection{Subspace Codes}

The following results are subsumed by the results of the
subsequent subsection, but we feel the presentation is enhanced by
deriving the subspace case first since it can be proved in a more
elementary fashion. We begin with a refinement of the
Knill-Laflamme Theorem that will be useful for our purposes.

\begin{lemma}\label{lem:coderep01}
    Let $\mathcal{E}:\mathcal{L}_1(\mathcal{H}) \rightarrow
    \mathcal{L}_1(\mathcal{H})$ be a quantum operation,
    and let $\mathcal{C} \subseteq \mathcal{H}$ be a subspace. Then $\mathcal{C}$
    is correctable for $\mathcal{E}$ if and only if there is a mixed unitary
    channel $\mathcal{F} \equiv \left\{ \sqrt{p_{i}} U_{i} \right\}$ such that
    $\mathcal{E}(\rho) = \mathcal{F}(\rho)$ for all $\rho \in \mathcal{L}_1(\mathcal{C})$
    and $P_{C}U_{i}^{\dagger}U_{j}P_{C} = 0$ for all $i \neq j$.
\end{lemma}

\begin{proof}
The code matrix $\Lambda = (\lambda_{ij})$ for $\mathcal{C}$ and
$\mathcal{E}\equiv \{E_j\}$ is  a density matrix, and thus there
is a unitary matrix $U = (u_{ij})$ such that $U \Lambda
U^{\dagger}$ is diagonal (call this diagonal matrix $D =
(d_{ij})$). Define a map $\mathcal{F} \equiv \left\{ F_{i}
\right\}$ where
\begin{align*}
 F_{i} = \sum_{j}{\overline{u_{ij}}E_{j}}.
\end{align*}
Note that $\mathcal{E} = \mathcal{F}$. Furthermore, for all $i,j,$
it is the case that
\begin{equation*}
P_{\mathcal{C}}F_{i}^{\dagger}F_{j}P_{\mathcal{C}}  = \sum_{k,l}{u_{ik}\overline{u_{jl}}}P_{\mathcal{C}}E_{k}^{\dagger}E_{l}P_{\mathcal{C}} \\
  = \sum_{k,l}{u_{ik}\overline{u_{jl}}}\lambda_{kl}P_{\mathcal{C}}
  = d_{ij}P_{\mathcal{C}}.
\end{equation*}
Thus $P_{\mathcal{C}}F_{i}^{\dagger}F_{j}P_{\mathcal{C}} = 0$ for
all $i \neq j$. For each $i$, we can apply the polar decomposition
to obtain unitary operators $U_i$ such that
\[
F_iP_{\mathcal{C}} = U_i
\sqrt{P_{\mathcal{C}}F_{i}^{\dagger}F_{i}P_{\mathcal{C}}} =
\sqrt{d_{ii}} \, U_i P_{\mathcal{C}}.
\]
When restricted to $\mathcal{L}_1(\mathcal{C})$, the mixed unitary
channel $\mathcal{F}^\prime \equiv \{ \sqrt{d_{ii}} U_i \}$ is
equivalent to the restriction of $\mathcal{F}$ (and hence
$\mathcal{E}$) to $\mathcal{L}_1(\mathcal{C})$, and has the
desired orthogonality property.
\end{proof}

To illustrate  Lemma~\ref{lem:coderep01} we introduce a simple
example.

\begin{exam}\label{exam:subspace}
{\rm Let $I$ be the $2 \times 2$ identity matrix, and let $U$ and
$V$ be $2 \times 2$ unitary matrices, let $q \in (0, 1)$, and let
$\mathcal{H}$ be two-qubit ($4$-dimensional) Hilbert space with
standard basis $\left\{ \ket{00}, \ket{01}, \ket{10}, \ket{11}
\right\}$. Then consider the channel $\mathcal{E}$ given by the
four Kraus operators represented in the standard basis as
\begin{align*}
\alpha\left[ \begin{matrix} I & U \\ 0 & 0 \end{matrix} \right],
\quad \alpha\left[ \begin{matrix} I & -U \\ 0 & 0 \end{matrix}
\right], \quad \beta\left[ \begin{matrix} I & V \\ I & V
\end{matrix} \right], \quad \beta\left[ \begin{matrix} -I & V \\ I & -V
\end{matrix} \right],
\end{align*}
where $\alpha = \frac{\sqrt{q}}{\sqrt{2}}$ and $\beta =
\frac{\sqrt{1 - q}}{2}$. It is easily verified that $C = \spn
\left\{ \ket{00}, \ket{01} \right\}$ is a correctable subspace for
$\mathcal{E}$ with projection $P_{C} = \ketbra{00}{00} +
\ketbra{01}{01}$.

Lemma~\ref{lem:coderep01} tells us then that there exists a mixed
unitary channel $\mathcal{F}$ such that
$\mathcal{E}|_{\mathcal{L}_1(\mathcal{C})} =
\mathcal{F}|_{\mathcal{L}_1(\mathcal{C})}$. Indeed, it is not
difficult to verify that
\[
\mathcal{F} = \left\{ \frac{\sqrt{1 + q}}{\sqrt{2}} I \otimes I,
\frac{\sqrt{1 - q}}{\sqrt{2}}X \otimes I \right\}
\]
is such a channel because for all $\rho\in\mathcal{L}_1({\mathbb
C}^2)$ we have
\[
\mathcal{E}(\ketbra{0}{0} \otimes \rho) =
\mathcal{F}(\ketbra{0}{0} \otimes \rho) = (\frac{1}{2}I +
\frac{q}{2}Z) \otimes \rho.
\]
}
\end{exam}

The following result shows that any quantum operation restricted
to a correctable code subspace can be described by a
representation, up to ``smearing'' by a fixed operator given by
the image of the code projection under the map.

\begin{thm}\label{thm:repsubspace}
Let $\mathcal{E}:\mathcal{L}_1(\mathcal{H}) \rightarrow
\mathcal{L}_1(\mathcal{H})$ be a quantum operation, and let
$\mathcal{C} \subseteq \mathcal{H}$ be a subspace. Then the
following are equivalent:
\begin{enumerate}
    \item[(i)] $\mathcal{C}$ is correctable for $\mathcal{E}$.
    \item[(ii)] There is a representation $\pi : \mathcal{L}_1
    (\mathcal{C}) \rightarrow \mathcal{L}_1(\mathcal{H})$ such that:
    \begin{equation*}\label{coderepn}
    \mathcal{E}( \rho ) = \pi( \rho ) \mathcal{E}( P_{\mathcal{C}} ) =
    \mathcal{E}( P_{\mathcal{C}} ) \pi( \rho ) \quad \forall \rho \in \mathcal{L}_1
    (\mathcal{C}).
    \end{equation*}
\end{enumerate}
Furthermore, $\pi^{\dagger}$ is a quantum operation that acts as a
correction operation for $\mathcal{E}$ on $\mathcal{C}$.
\end{thm}
\begin{proof}
We first prove the implication (1) $\Rightarrow$ (2). Since
$\mathcal{C}$ is correctable for $\mathcal{E}$, we know by
Lemma~\ref{lem:coderep01} that there exists a mixed unitary
channel $\mathcal{F} = \left\{ \sqrt{p_{i}}U_{i} \right\}$ such
that $\mathcal{F}( \rho ) = \mathcal{E}( \rho )$ for all $\rho \in
\mathcal{L}_1( \mathcal{C} )$ and $P_{C}U_{i}^{\dagger}U_{j}P_{C}
= 0$ whenever $i \neq j$. Define partial isometries $V_i =
U_{i}P_{C}$. It follows that the map  $\pi :
\mathcal{L}_1(\mathcal{C}) \rightarrow \mathcal{L}_1(\mathcal{H})$
defined by $\pi( \rho ) = \sum_{j}{V_{j}\rho V_{j}^{\dagger}}$ is
a $*$-homomorphism. Since the $V_j$ have mutually orthogonal
ranges, we have $\sum_j V_j V_j^\dagger \leq I$, and thus the map
$\pi^\dagger \equiv \{ V_j^\dagger \}$ is trace non-increasing.
(We can assume with no loss of generality that $\pi^\dagger$ is
trace preserving by including the projection onto the orthogonal
complement of the ranges of the $V_j$.) We further have for all
$\rho \in \mathcal{L}_1 (\mathcal{C})$,
\begin{eqnarray*}
  \mathcal{E}(P_{\mathcal{C}})\pi(\rho)  = \sum_{i,j}{p_{i}V_{i}V_{i}^{\dagger}V_{j}\rho
  V_{j}^{\dagger}}
   = \sum_{i}{p_{i}V_{i} \rho V_{i}^{\dagger}}
   = \sum_{i}{p_{i}U_{i} \rho U_{i}^{\dagger}}
   = \mathcal{E}( \rho ).
\end{eqnarray*}
A similar argument shows that $\mathcal{E}( \rho ) =
\pi(\rho)\mathcal{E}(P_{\mathcal{C}})$.

To see (2) $\Rightarrow$ (1), observe that the equation
$\mathcal{E}( \rho ) = \pi(\rho)\mathcal{E}(P_{\mathcal{C}})$ and
trace preservation of $\mathcal{E}$ implies
\begin{equation*}
 \Tr\left( \rho \right) =  \Tr\left( \mathcal{E}( \rho ) \right) = \Tr\left( \pi(\rho)\mathcal{E}(P_{\mathcal{C}}) \right)
  = \Tr\left( \rho\pi^{\dagger}(\mathcal{E}(P_{\mathcal{C}})) \right).
\end{equation*}
Since this equation holds for all $\rho \in
\mathcal{L}_1(\mathcal{C})$, we have $P_\mathcal{C} =
P_\mathcal{C} \pi^\dagger(\mathcal{E}(P_\mathcal{C}))
P_\mathcal{C}$, and hence by trace preservation of
$\pi^\dagger\circ\mathcal{E}$ that
\begin{align}\label{eq:sspc_1}
  P_{\mathcal{C}} = \pi^{\dagger}(\mathcal{E}(P_{\mathcal{C}})).
\end{align}
Note that $\Tr(\pi^{\dagger}(\alpha)\beta\gamma) =
\Tr(\alpha\pi(\beta\gamma)) = \Tr(\alpha\pi(\beta)\pi(\gamma)) =
\Tr(\pi^{\dagger}(\alpha\pi(\beta))\gamma)$ for all $\alpha,
\beta, \gamma \in \mathcal{L}_1(\mathcal{H})$. Since this equation
holds for all $\gamma \in \mathcal{L}_1(\mathcal{H})$ in
particular, we have that:
\begin{align}\label{eq:sspc_2}
\pi^{\dagger}(\alpha)\beta = \pi^{\dagger}(\alpha\pi(\beta))
\quad\quad \forall \alpha,\beta \in \mathcal{L}_1(\mathcal{H}).
\end{align}
Multiplying Eq.~\eqref{eq:sspc_1} on the right by an arbitrary
$\rho \in \mathcal{L}_1(\mathcal{C})$ now shows that $\rho =
\pi^{\dagger}(\mathcal{E}(P_{\mathcal{C}}))\rho$. If we then apply
Eq.~\eqref{eq:sspc_2} with $\alpha = \mathcal{E}(P_{\mathcal{C}})$
and $\beta = \rho$, we see that
\[
\rho = \pi^{\dagger}(\mathcal{E}(P_{\mathcal{C}}))\rho =
\pi^{\dagger}(\mathcal{E}(P_{\mathcal{C}})\pi(\rho)) =
\pi^{\dagger}(\mathcal{E}(\rho)),
\]
and this completes the proof.
\end{proof}

Observe from the above proof that if $\mathcal{F} = \left\{
\sqrt{p_{i}}U_{i} \right\}$ is the mixed unitary channel described
by Lemma~\ref{lem:coderep01}, then the representation described by
Theorem~\ref{thm:repsubspace} is given by $\pi(\rho) =
\sum_{i}{V_{i}\rho V_{i}^{\dagger}}$, where $V_{i} =
U_{i}P_{\mathcal{C}}$. Similarly, the correction operation is
given by $\pi^{\dagger}(\sigma) = \sum_{i}{V_{i}^{\dagger}\sigma
V_{i}}$.

\begin{exam}
{\rm Returning to Example~\ref{exam:subspace}, we see that
\begin{align*}
\pi(\rho) & = \left[\begin{matrix} I & 0 \\ 0 & 0 \end{matrix}\right] \rho \left[\begin{matrix} I & 0 \\ 0 & 0 \end{matrix}\right] + \left[\begin{matrix} 0 & 0 \\ I & 0 \end{matrix}\right] \rho \left[\begin{matrix} 0 & I \\ 0 & 0 \end{matrix}\right] \\
\end{align*}
and
\begin{align*}
\pi^{\dagger}(\sigma) & = \left[\begin{matrix} I & 0 \\ 0 & 0
\end{matrix}\right] \sigma \left[\begin{matrix} I & 0 \\ 0 & 0
\end{matrix}\right] + \left[\begin{matrix} 0 & I \\ 0 & 0
\end{matrix}\right] \sigma \left[\begin{matrix} 0 & 0 \\ I & 0
\end{matrix}\right].
\end{align*}
Note that $\pi^{\dagger}$ is indeed  a correction operation for
this channel on the subspace $\mathcal{C}$ because for all $\rho
\in \mathcal{L}_1({\mathbb C}^2)$
\[
\pi^{\dagger} \circ \mathcal{E} (\ketbra{0}{0} \otimes \rho) =
\pi^{\dagger}\Big((\frac{1}{2}I + \frac{q}{2}Z) \otimes \rho\Big)
= \ketbra{0}{0} \otimes \rho .
\]
}
\end{exam}

\vspace{0.1in}

\subsection{Subsystem Codes}

We next extend the results of the previous subsection to the more
general case of subsystem codes. We begin with a pair of technical
results, firstly the direct generalization of
Lemma~\ref{lem:coderep01} for subsystem codes. This result
formalizes a key component of the proof of the main result from
\cite{KS06}. Recall we are using the notation $\fA_\mathcal{B}:=
1_\mathcal{A}\otimes\mathcal{L}_1(\mathcal{B})$.

\begin{lemma}\label{lem:coderep02}
    Let $\mathcal{E}:\mathcal{L}_1(\mathcal{H}) \rightarrow \mathcal{L}_1(\mathcal{H})$
    be a quantum operation, and let $\mathcal{C} = \mathcal{A}\otimes\mathcal{B} \subseteq \mathcal{H}$
    be a subspace. Then $\mathcal{B}$ is correctable for $\mathcal{E}$ if and only if there is a
    channel $\mathcal{G}$ with $\mathcal{G}\circ\mathcal{P}_\mathcal{C} \equiv \left\{ V_{i}(D_{i}
    \otimes I_{\mathcal{B}}) \right\}$ such that
    $\mathcal{E}(\rho) = \mathcal{G}(\rho)$ for all $\rho \in
    \fA_{\mathcal{B}}$, where $V_i$ are unitary operators, $D_{i}$ are mutually commuting positive operators,
    and
    $P_{\mathcal{C}_i}V_{i}^{\dagger}V_{j}P_{\mathcal{C}_j} = \delta_{ij}P_{\mathcal{C}_i}$ for all $i,j$,
    where $\mathcal{C}_i = {\rm Ran}\,(D_i)\otimes\mathcal{B}\subseteq\mathcal{C}$.
\end{lemma}

\begin{proof}
If there is such a channel $\mathcal{G}$, then it is easily
verified that the channel $\mathcal{R} \equiv \{ V_i^\dagger
P_{\mathcal{C}_i} \}$ acts as a $\mathcal{B}$ subsystem recovery
operation for $\mathcal{E}$. For the other direction, begin by
noting that if $\mathcal{B}$ is correctable for $\mathcal{E}$,
then there exist operators $F_{ij}$ on $\mathcal{A}$ such that
\begin{align}\label{subsys_01}
P_{\mathcal{C}}E_{i}^{\dagger}E_{j}P_{\mathcal{C}} = F_{ij}
\otimes I_{\mathcal{B}} \quad\quad \forall i,j.
\end{align}
Observe that the operator block matrix $F = (F_{ij})$ is positive
since
\[
(I_{m} \otimes P_{\mathcal{C}}) E^{\dagger}E (I_{m}\otimes
P_{\mathcal{C}}) = F \otimes I_{\mathcal{B}},
\]
where the row matrix $E = [E_{1} E_{2} \cdots E_{m}]$, the number
of $E_{i}$ is $m$, and $I_{m}$ is the identity operator on
$m$-dimensional Hilbert space. Assume that we have a matrix
representation for each of the $F_{ij}$, and hence for
$F=(F_{ij})$, defined by a fixed basis for $\mathcal{A}$. Thus we
let $U$ be a unitary matrix such that $UFU^{\dagger} = D$ is
diagonal and let $U = (U_{ij})$ and $D = (D_{ij})$ be the
associated block decompositions. We may naturally regard each
$U_{ij}$ as the matrix representation (in the fixed basis) for an
operator on $\mathcal{A}$. Then
\begin{align}\label{subsys_02}
\sum_{k,l}{U_{ik}F_{kl}U_{jl}^{\dagger}} = \delta_{ij}D_{ii} \quad\quad \forall i,j,\\
\label{subsys_03} \sum_{k}{U_{ki}^{\dagger}U_{kj}} =
\delta_{ij}I_{\mathcal{A}} \quad\quad \forall i,j.
\end{align}

Next define a channel $\mathcal{G} \equiv \left\{G_{i}\right\}$
where for all $i$,
\begin{align*}
G_{i} = \sum_{j}{E_{j}(U_{ij}^{\dagger}\otimes
I_{\mathcal{B}})P_{\mathcal{C}} + E_{i}P_{\mathcal{C}}^{\perp}}.
\end{align*}
Let $X_{ij} = E_{j}(U_{ij}^{\dagger} \otimes
I_{\mathcal{\mathcal{B}}})P_{\mathcal{\mathcal{C}}}$. Then by
Eqs.~\eqref{subsys_01} and \eqref{subsys_02}, one can verify that
for all $i,j$,
\begin{align*}
P_{\mathcal{C}}G_{i}^{\dagger}G_{j}P_{\mathcal{C}}  =
\sum_{k,l}{X_{ik}^{\dagger}X_{jl}}
  = \Big( \sum_{k,l}{U_{ik}F_{kl}U_{jl}^{\dagger}} \Big) \otimes I_{\mathcal{B}}
  = D_{ij} \otimes I_{\mathcal{B}},
\end{align*}
and $D_{ij} = 0$ for all $i \neq j$. Moreover,
Eq.~\eqref{subsys_03} yields for all
$I_\mathcal{A}\otimes\rho_{\mathcal{B}}\in\fA_\mathcal{B}$
\begin{align*}
\mathcal{G}(I_{\mathcal{A}}\otimes \rho_{\mathcal{B}}) & = \sum_{i}{G_{i}(I_{\mathcal{A}} \otimes \rho_{\mathcal{B}})G_{i}^{\dagger}} \\
 & = \sum_{i,j,k}{X_{ij}(I_{\mathcal{A}}\otimes \rho_{\mathcal{B}})X_{ik}^{\dagger}} \\
 & = \sum_{j,k}{E_{j}(\left(\sum_{i}{U_{ij}^{\dagger}U_{ik}}\right) \otimes \rho_{\mathcal{B}})E_{k}^{\dagger}} \\
 & = \sum_{j}{E_{j}(I_{\mathcal{A}}\otimes \rho_{\mathcal{B}})E_{j}^{\dagger}} \\
 & = \mathcal{E}(I_{\mathcal{A}}\otimes \rho_{\mathcal{B}}).
\end{align*}
By the polar decomposition applied to each $G_{i}P_{\mathcal{C}}$,
and the fact that these operators have mutually orthogonal ranges,
there are unitaries $V_{i}$  such that
\begin{align*}
G_{i}P_{\mathcal{C}} =
V_{i}\sqrt{P_{\mathcal{C}}G_{i}^{\dagger}G_{i}P_{\mathcal{C}}} =
V_{i}\left(\sqrt{D_{ii}} \otimes I_{\mathcal{B}}\right).
\end{align*}
Let $D_{i} = \sqrt{D_{ii}}$ and let $\mathcal{C}_i = {\rm
Ran}\,(D_i)\otimes \mathcal{B}$. Observe that each partial
isometry $V_i P_\mathcal{C}$ has $\mathcal{C}_i$ as its initial
projection and that the final projections are onto mutually
orthogonal subspaces. Hence we have $P_{\mathcal{C}_i} V_i^\dagger
V_j P_{\mathcal{C}_j} = \delta_{ij} P_{\mathcal{C}_i}$. Thus any
channel $\mathcal{G}^\prime$ with $\mathcal{G}^\prime \circ
\mathcal{P}_\mathcal{C}\equiv \{V_i(D_i\otimes I_\mathcal{B})\}$
has the desired properties, up to the mutually commuting
condition. However, observe that each $D_i$ can be replaced by
$U_i D_i U_i^\dagger$, where $U_i$ is an arbitrary unitary
operator on $\mathcal{A}$, without affecting the result. Thus, we
can arrange things so that the $D_i$ are simultaneously
diagonalizable and commute.
\end{proof}

This is all we need to prove Theorem~\ref{thm:repsubsys}. However,
notice that the preceding result shows what the map $\mathcal{E}$
looks like when restricted to the algebra $\fA_\mathcal{B}$, but
it is not clear how, or even if, this extends to the entire
subspace $\mathcal{C}$. We extend this result as follows.

\begin{thm}\label{thm:coderep03}
Let $\mathcal{E}:\mathcal{L}_1(\mathcal{H}) \rightarrow
\mathcal{L}_1(\mathcal{H})$ be a quantum operation, and let
$\mathcal{C} = \mathcal{A} \otimes \mathcal{B} \subseteq
\mathcal{H}$ be a subspace. Then $\mathcal{B}$ is correctable for
$\mathcal{E}$ if and only if there is a family of unitary
operators $\big\{ U_{i} \big\}$ with
$P_{\mathcal{C}}U_{i}^{\dagger}U_{j}P_{\mathcal{C}} = 0$ for all
$i \neq j$ and a channel $\mathcal{N}_{\mathcal{A}} :
\mathcal{L}_1(\mathcal{A}) \rightarrow \mathcal{L}_1(\mathcal{A})$
with Kraus operators $\big\{ N_{i,j} \big\}$ such that
$\mathcal{E}(\rho) = \mathcal{F}(\rho)$ for all $\rho \in
\mathcal{L}_1(\mathcal{C})$, where $\mathcal{F} :
\mathcal{L}_1(\mathcal{H}) \rightarrow \mathcal{L}_1(\mathcal{H})$
is the channel given by the Kraus operators $\big\{ U_{i}(N_{i,j}
\otimes I_{\mathcal{B}}) \big\}$.
\end{thm}

\begin{proof}
First let $\ket{\psi} \in B$ be a unit vector and set $P =
\ketbra{\psi}{\psi}$. Suppose that $\big\{ \ket{\alpha_k} \big\}$
is an orthonormal basis for $\mathcal{A}$ and set $A_{k} =
\ketbra{\alpha_k}{\alpha_k}$. Now define $Q_{i} =
U_{i}(I_{\mathcal{A}} \otimes P)U_{i}^{\dagger}$, where $\big\{
U_{i} \big\}$ is the family of unitary operators given by
Lemma~\ref{lem:coderep02}. Note that each $Q_i$ is an orthogonal
projection. Furthermore, it is not difficult to verify that
\begin{align*}
  0 \leq \sum_{i}Q_{i} \mathcal{E}(A_k \otimes P) Q_{i} \leq
  \mathcal{E}(A_k \otimes P) \leq \mathcal{E}(I_{\mathcal{A}} \otimes P)
  = \sum_{i}U_{i}(D_{i}^{2} \otimes P)U_{i}^{\dagger},
\end{align*}
where $\big\{ D_i \big\}$ is the family of positive diagonal
operators given by Lemma~\ref{lem:coderep02}. Since the above
inequalities hold for all $k$ and
\begin{align*}
  \mathcal{E}(I_{\mathcal{A}} \otimes P) = \sum_{k}\mathcal{E}(A_{k}
  \otimes P) = \sum_{i,k} Q_{i}\mathcal{E}(A_{k} \otimes P)Q_{i},
\end{align*}
it follows that $\sum_{i} Q_{i} \mathcal{E}(A_{k} \otimes P) Q_{i}
= \mathcal{E}(A_{k} \otimes P)$ for all $k$. A simple
dimension-counting argument then shows that $\mathcal{E}(A_{k}
\otimes P)$ must be of the form
\begin{align*}
  \mathcal{E}(A_k \otimes P) = \sum_{i}U_{i}(\sigma_{i,k,\psi} \otimes P)U_{i}^{\dagger}.
\end{align*}
It can also be shown via a standard linearity argument that the
operators $\big\{ \sigma_{i,k,\psi} \big\}$ do not depend on
$\ket{\psi}$. Thus, it follows from linearity of $\mathcal{E}$
that for all $\sigma_{\mathcal{A}}$ there exist positive operators
$\big\{ \tau_{{\mathcal{A}},i} \big\}$ such that
\begin{align*}
  \mathcal{E}(\sigma_{\mathcal{A}} \otimes \rho_{\mathcal{B}}) =
  \sum_{i}U_{i}(\tau_{{\mathcal{A}},i} \otimes \rho_{\mathcal{B}})U_{i}^{\dagger}
  \quad\quad \forall \, \rho_{\mathcal{B}}.
\end{align*}
The proof is completed by defining
$\mathcal{N}_{\mathcal{A}}(\sigma_\mathcal{A}) = \sum_i
\tau_{\mathcal{A},i}$.
\end{proof}

The following description of subsystem codes in the
Schr\"{o}dinger picture complements other descriptions such as
those found in \cite{KS06,Kni06,BKP08,BKK08}.


%

\begin{thm}\label{thm:repsubsys}
Let $\mathcal{E}:\mathcal{L}_1(\mathcal{H}) \rightarrow
\mathcal{L}_1(\mathcal{H})$, and let $\mathcal{C} = \mathcal{A}
\otimes \mathcal{B} \subseteq \mathcal{H}$ be a subspace. Then the
following are equivalent:
\begin{enumerate}
    \item $\mathcal{B}$ is a correctable subsystem for $\mathcal{E}$.
    \item There is a representation $\pi : \mathfrak{A}_\mathcal{B} \rightarrow
    \mathcal{L}_1(\mathcal{H})$ such that
    \[
    \mathcal{E}( \rho ) = \pi( \rho ) \mathcal{E}( P_{\mathcal{C}} ) = \mathcal{E}( P_{\mathcal{C}} )
    \pi( \rho ) \quad \quad \forall \rho \in \mathfrak{A}_\mathcal{B}.
    \]
\end{enumerate}
\end{thm}
\begin{proof}

To prove the implication (1) $\Rightarrow$ (2), note that since
$\mathcal{B}$ is correctable for $\mathcal{E}$, we know by
Lemma~\ref{lem:coderep02} that there exists a channel
$\mathcal{G}$ with $\mathcal{G}\circ\mathcal{P}_\mathcal{C} \equiv
\left\{V_{i}(D_{i}\otimes I_{\mathcal{B}})\right\}$ such that
$\mathcal{G}( I_{\mathcal{A}}\otimes\rho_{\mathcal{B}} ) =
\mathcal{E}( I_{\mathcal{A}}\otimes\rho_{\mathcal{B}} )$ for all
$\rho_{\mathcal{B}}$, and $\left\{V_{i}\right\}$ is a family of
partial isometries such that $V_{i}^{\dagger}V_{j} = 0$ whenever
$i \neq j$ and $V_{i}^{\dagger}V_{i} = P_{\mathcal{C}_i}$, where
$P_{\mathcal{C}_i}$ is the orthogonal projection onto
$\mathcal{C}_i = {\rm Ran}\,(D_i)\otimes\mathcal{B}$.

Now define $\pi : \fA_\mathcal{B} \rightarrow
\mathcal{L}_1(\mathcal{H})$ by $\pi( I_{\mathcal{A}}\otimes
\rho_{\mathcal{B}} ) =
\sum_{i}{V_{i}(I_{\mathcal{A}}\otimes\rho_{\mathcal{B}})V_{i}^{\dagger}}$.
Then $\pi$ is easily seen to be a  $*$-homomorphism on
$\fA_\mathcal{B}$ (using the fact that
$P_{\mathcal{C}_i}=Q_i\otimes I_\mathcal{B}$ for some projection
$Q_i$ on $\mathcal{A}$). Its dual $\pi^\dagger = \{ V_i^\dagger
\}$ is trace non-increasing and can be trivially extended to a
trace preserving map as before.  It then follows that
\begin{align*}
  \mathcal{E}(P_{\mathcal{C}})\pi(I_{\mathcal{A}}\otimes\rho_{\mathcal{B}}) & =
  \Big(\sum_{i}{V_{i}(D_{i}\otimes I_{\mathcal{B}})P_{\mathcal{C}}(D_{i}^{\dagger}\otimes
  I_{\mathcal{B}})V_{i}^{\dagger}}\Big)\Big(\sum_{j}{V_{j}(I_{\mathcal{A}}\otimes\rho_{\mathcal{B}})
  V_{j}^{\dagger}}\Big) \\
  & = \sum_{i}{V_{i}(D_{i}\otimes I_{\mathcal{B}})P_{\mathcal{C}}(D_{i}^{\dagger}\otimes I_{\mathcal{B}}) P_{\mathcal{C}_i} (I_{\mathcal{A}}\otimes\rho_{\mathcal{B}}) V_{i}^{\dagger}} \\
  & = \sum_{i}{V_{i}(D_{i}\otimes I_{\mathcal{B}}) (I_{A}\otimes\rho_{\mathcal{B}}) (D_{i}^{\dagger}\otimes I_{\mathcal{B}}) V_{i}^{\dagger}} \\
  & = \mathcal{G}( I_{\mathcal{A}}\otimes\rho_{\mathcal{B}} )  = \mathcal{E}( I_{\mathcal{A}}\otimes\rho_{\mathcal{B}} ).
\end{align*}
A similar argument shows that $\mathcal{E}(
I_{\mathcal{A}}\otimes\rho_{\mathcal{B}} ) =
\pi(I_{\mathcal{A}}\otimes\rho_{\mathcal{B}})\mathcal{E}(P_{\mathcal{C}})$.

To see (2) $\Rightarrow$ (1), we show that the algebra
$\fA_\mathcal{B}$ may be precisely corrected, which is equivalent
to correcting the subsystem $\mathcal{B}$ (see Theorem~3.2 of
\cite{KLPL06} for instance). First note that the representation
$\pi$ defines a subspace and subsystems $\mathcal{C}' =
\mathcal{A}' \otimes \mathcal{B}'$ with $\mathcal{B}'$ the same
dimension as $\mathcal{B}$ and an isometry $V: \mathcal{B}
\rightarrow \mathcal{B}'$ such that
\[
\pi(I_\mathcal{A}\otimes \rho_\mathcal{B}) = I_{\mathcal{A}'}
\otimes \mathcal{V} (\rho_\mathcal{B}) \quad \forall
\rho_\mathcal{B},
\]
where $\mathcal{V} (\rho_\mathcal{B}) = V \rho_\mathcal{B}
V^\dagger$. Further, as $\mathcal{E} (P_\mathcal{C})$ commutes
with $\pi(\fA_\mathcal{B})$, it follows that $P_{\mathcal{C}'}
\mathcal{E} (P_\mathcal{C}) P_{\mathcal{C}'} =
\sigma_{\mathcal{A}'} \otimes I_{\mathcal{B}'}$ for some positive
operator $\sigma_{\mathcal{A}'}\in\mathcal{L}(\mathcal{A}')$ with
trace equal to $\dim\mathcal{C}$. Thus we have for all
$\rho_\mathcal{B}$,
\begin{eqnarray*}
\mathcal{E}(I_\mathcal{A}\otimes \rho_\mathcal{B}) &=&
\pi(I_\mathcal{A}\otimes
\rho_\mathcal{B})\mathcal{E}(P_\mathcal{C}) \\ &=&
(I_{\mathcal{A}'}\otimes
\mathcal{V}(\rho_\mathcal{B}))(\sigma_{\mathcal{A}'}\otimes
I_{\mathcal{B}'}) \\ &=& \sigma_{\mathcal{A}'} \otimes
\mathcal{V}(\rho_\mathcal{B}).
\end{eqnarray*}
Now define a channel $\mathcal{R}$ on $\mathcal{H}$ such that
$\mathcal{R} \circ \mathcal{P}_{\mathcal{C}'} =
(\mathcal{D}_{\mathcal{A}|\mathcal{A}'} \otimes
\mathcal{V}^\dagger) \circ \mathcal{P}_{\mathcal{C}'}$, where
$\mathcal{D}_{\mathcal{A}|\mathcal{A}'}$ is the complete
depolarizing channel from $\mathcal{A}'$ to $\mathcal{A}$, and it
follows that
$(\mathcal{R}\circ\mathcal{E})(I_\mathcal{A}\otimes\rho_\mathcal{B})
= I_\mathcal{A} \otimes \rho_\mathcal{B}$ for all
$\rho_\mathcal{B}$. This shows $\fA_\mathcal{B}$ can be exactly
corrected, and completes the proof.

\end{proof}

\section{The Multiplicative Domain and Unitarily Correctable Codes}

Given a CP map $\phi:
\fA \rightarrow \fB$ between two operator algebras, the
\emph{multiplicative domain} of $\phi$, denoted $MD(\phi)$, is
effectively the largest subalgebra of $\fA$ for which the
restriction of $\phi$ is a multiplicative map. It is explicitly
defined as follows:
\[
MD(\phi) :=  \big\{ a \in \fA : \phi(a)\phi(b) = \phi(ab)\text{
and } \phi(b)\phi(a) = \phi(ba) \text{ for all } b \in \fA \big\}.
\]
It is clear that $MD(\phi)$ is an algebra, and hence has a
structure as in Eq.~(\ref{opalgform}). In this section we address
this basic question: What role, if any, does the multiplicative
domain play in quantum error correction?

The unital case ($\phi(I)=I$) often stands out in the CP theory,
and this is the case for multiplicative domains. The following
result of the first named author \cite{Cho74,Paulsentext} shows
how the multiplicative domain simplifies in the unital case. Note
that in particular, if $\mathcal{E}$ is a quantum channel then
Theorem~\ref{thm:multDom1} applies to $\mathcal{E}^{\dagger}$.

\begin{thm}\label{thm:multDom1}
  Let $\fA$ and $\fB$ be algebras and let $\phi : \mathcal{A} \mapsto \mathcal{B}$ be a completely positive, unital map. Then
  \begin{align*}
   MD(\phi) = & \big\{ a \in \mathcal{A} : \phi(a)^{\dagger}\phi(a) =
\phi(a^{\dagger}a)\text{ and } \phi(a)\phi(a)^{\dagger} =
\phi(aa^{\dagger})\big\}.
  \end{align*}
  Furthermore, $\phi$ is a $*$-homomorphism when restricted to this set.
\end{thm}

Turning to quantum error correction, an important class of quantum
codes are the so-called ``unitarily correctable codes'' (UCC).
These are codes for which a unitary recovery operation can be
obtained. Alternatively, UCCs are the highly degenerate codes for
which a recovery operation can be implemented without a
measurement. As such, they are potentially quite useful in fault
tolerant quantum computing since these codes and their recovery
operations do not require more of the system Hilbert space than
what is required by the initial code. A subsystem code
$\mathcal{B}$ is \emph{unitarily correctable} for $\mathcal{E}$ if
there is a unitary operation $\mathcal{U}$ and channel
$\mathcal{F}_\mathcal{A}: \mathcal{L}_1(\mathcal{A}) \rightarrow
\mathcal{L}_1(\mathcal{A})$ such that
\[
\mathcal{E} \circ \mathcal{P}_{\mathcal{AB}} = \mathcal{U} \circ
(\mathcal{F}_\mathcal{A}\otimes id_\mathcal{B}) \circ
\mathcal{P}_\mathcal{AB}.
\]
The UCC class includes decoherence-free subspaces and noiseless
subsystems in the case that $\mathcal{U}=id$.

The results of the previous section motivate a new notion for
codes in which UCC stand out as a special case.

\begin{defn}
{\rm Let
$\mathcal{C}=\mathcal{A}\otimes\mathcal{B}\subseteq\mathcal{H}$,
and suppose $\mathcal{B}$ is correctable for
$\mathcal{E}:\mathcal{L}_1(\mathcal{H})\rightarrow\mathcal{L}_1(\mathcal{H})$.
Then we define the \emph{correction rank} of $\mathcal{B}$ for
$\mathcal{E}$ to be the multiplicity of the representation $\pi$
determined by $\mathcal{E}$ and $\mathcal{B}$ as in
Theorem~\ref{thm:repsubsys}. }
\end{defn}

Observe that in the case of subspace codes the UCC for a given
channel $\mathcal{E}$ are precisely its correction rank-1 codes.


One of the main results from \cite{KS06} shows in the unital case
($\mathcal{E}(I)=I$) that UCCs are precisely the passive codes for
the map composed with its dual.

\begin{thm}\label{thm:kribSpek1}
  \cite{KS06} Let $\mathcal{E}$ be a unital quantum operation. Then the following are equivalent:
  \begin{enumerate}
    \item $\mathcal{B}$ is a unitarily correctable subsystem for $\mathcal{E}$.
    \item $\mathcal{B}$ is a noiseless subsystem for $\mathcal{E}^{\dagger}\circ\mathcal{E}$.
  \end{enumerate}
\end{thm}

Theorem~\ref{thm:kribSpek1} shows that we may unambiguously define
\emph{the} UCC algebra for a unital channel
$\mathcal{E}\equiv\{E_i\}$ as
\[
UCC(\mathcal{E}):= \{ \rho :
\mathcal{E}^\dagger\circ\mathcal{E}(\rho) = \rho\} =\{\rho :
[\rho,E_i^\dagger E_j]=0\},
\]
as we know from the theory of passive quantum error correction
that the latter algebra encodes all noiseless subsystems for
$\mathcal{E}^\dagger\circ\mathcal{E}$. (See \cite{KS06} and
references therein for further discussions on this point.)

The following theorem shows the intimate relationship between a
unital channel's unitarily correctable codes, its multiplicative
domain, and the unitarily correctable codes and multiplicative
domain of its dual map. Interestingly, in the case of a unital
channel this shows that a naturally arising object in the theory
of CP maps, the multiplicative domain, describes a class of
quantum codes that have arisen in quantum error correction for
completely different reasons.

\begin{thm}\label{thm:multDom2}
  Let $\mathcal{E}$ be a unital quantum operation. Then the following four algebras coincide:
  \begin{enumerate}
    \item $MD(\mathcal{E})$
    \item $UCC(\mathcal{E})$
    \item $\mathcal{E}^\dagger(MD(\mathcal{E}^\dagger))$
    \item $\mathcal{E}^\dagger(UCC(\mathcal{E}^\dagger))$.
  \end{enumerate}
\end{thm}
\begin{proof}
As $\mathcal{E}$ is a unital channel if and only if
$\mathcal{E}^\dagger$ is the same, this result is symmetric in
$\mathcal{E}$ and $\mathcal{E}^\dagger$. We first show that
$MD(\mathcal{E}^{\dagger}) \subseteq UCC(\mathcal{E}^{\dagger})$.
Note that if $a \in MD(\mathcal{E}^{\dagger})$ then
$\Tr(\mathcal{E}^{\dagger}(a)\mathcal{E}^{\dagger}(b)) =
\Tr(\mathcal{E}^{\dagger}(ab))$ for all $b \in
\mathcal{L}_1(\mathcal{H})$. Then
$\Tr(\mathcal{E}\circ\mathcal{E}^{\dagger}(a)b) =
\Tr(\mathcal{E}(1)ab) = \Tr(ab)$ for all $b \in
\mathcal{L}_1(\mathcal{H})$ and so it follows that
$\mathcal{E}\circ\mathcal{E}^{\dagger}(a) = a$ for all $a \in
MD(\mathcal{E}^{\dagger})$. The inclusion then follows from
Theorem~\ref{thm:kribSpek1}.

To see the opposite inclusion, note that if $\mathcal{B}$ is a
unitarily correctable subsystem for $\mathcal{E}^{\dagger}$ then
Lemma~\ref{lem:coderep02} says that
$\mathcal{E}^{\dagger}\circ\mathcal{P}_{\mathcal{C}} \equiv \big\{
U(D \otimes I_{\mathcal{B}})P_{\mathcal{C}} \big\}$ for some
unitary $U$ and diagonal operator $D$. In fact, since
$\mathcal{B}$ is noiseless for the unital channel
$\mathcal{U}^\dagger\circ\mathcal{E}^\dagger$, it follows that
$\mathcal{U}^\dagger\circ\mathcal{E}^\dagger(I_{\mathcal{A}}
\otimes \rho_{\mathcal{B}}) = I_{\mathcal{A}} \otimes
\rho_{\mathcal{B}}$ for all $\rho_{\mathcal{B}}$. Hence we have $D
= I_\mathcal{A}$, and so $\mathcal{E}^\dagger(a) = \mathcal{U}(a)$
for all $a\in\fA_\mathcal{B}$. Theorem~\ref{thm:multDom1} now
shows the algebra $\fA_\mathcal{B}$, and hence
$UCC(\mathcal{E}^\dagger)$, is contained inside
$MD(\mathcal{E}^\dagger)$. Thus $MD(\mathcal{E}^{\dagger}) =
UCC(\mathcal{E}^{\dagger})$ (and similarly
$\mathcal{E}(MD(\mathcal{E})) = \mathcal{E}(UCC(\mathcal{E}))$).

We next show that $\mathcal{E}(UCC(\mathcal{E})) \subseteq
MD(\mathcal{E}^{\dagger})$. Now Theorem~\ref{thm:kribSpek1} says
that if $\mathcal{B}$ is unitarily correctable for $\mathcal{E}$
then $\mathcal{B}$ is noiseless for the unital channel
$\mathcal{E}^{\dagger}\circ\mathcal{E}$. Moreover, the restriction
of $\mathcal{E}$ to $\fA_\mathcal{B}$ is multiplicative by the
previous paragraph. Hence it follows that the restricted map
satisfies
$\mathcal{E}^{\dagger}\circ\mathcal{E}|_{\fA_\mathcal{B}} =
\mathcal{P}_{\mathcal{C}}|_{\fA_\mathcal{B}}$, and that
$\mathcal{E}^\dagger$ is a multiplicative map when restricted to
the image algebra $\mathcal{E}(\fA_\mathcal{B})$. Therefore from
Theorem~\ref{thm:multDom1} we have
$\mathcal{E}(\fA_\mathcal{B})\subseteq MD(\mathcal{E}^\dagger)$,
and the inclusion follows.

To get the opposite inclusion, note that $
\mathcal{E}^{\dagger}(UCC(\mathcal{E}^{\dagger}))  \subseteq
MD(\mathcal{E})$ implies
\begin{align*}
MD(\mathcal{E}^{\dagger}) = UCC(\mathcal{E}^{\dagger}) =
\mathcal{E} \circ
\mathcal{E}^{\dagger}(UCC(\mathcal{E}^{\dagger})) & \subseteq
\mathcal{E}(MD(\mathcal{E})) = \mathcal{E}(UCC(\mathcal{E})).
\end{align*}
The second equality above comes from Theorem~\ref{thm:kribSpek1}.
This completes the proof.
\end{proof}

Note that the equivalence of algebras $MD(\mathcal{E}^{\dagger})$
and $\mathcal{E}(UCC(\mathcal{E}))$ in Theorem~\ref{thm:multDom2}
does not imply that correctable codes that are not unitarily
correctable can not be found in the multiplicative domain of
$\mathcal{E}^{\dagger}$. The following example highlights this
fact, and presents a map that has a non-unitarily correctable code
with image under $\mathcal{E}$ that coincides with the image of a
unitarily correctable subsystem.

\begin{exam}\label{exam:multDom2}
{\rm Let $U,V,W \in \mathcal{L}(\mathcal{H})$ be unitary
operators, let $q \in [0, 1]$, and define a quantum channel
$\mathcal{E} : M_{2}(\mathcal{L}(\mathcal{H})) \mapsto
M_{2}(\mathcal{L}(\mathcal{H}))$ by the following pair of Kraus
operators:
\begin{align*}
E_{1} = q\begin{bmatrix} U & 0 \\ 0 & V \end{bmatrix} \ \ \ \ E_{2} = \sqrt{1 - q^2} \begin{bmatrix} 0 & U \\ W & 0 \end{bmatrix}.
\end{align*}

Then $\mathcal{E}$ is a unital quantum channel and a correctable
subspace for $\mathcal{E}$ is projected onto by the projection
\begin{align*}
  P_{\mathcal{C}} = \begin{bmatrix} I_{\mathcal{H}} & 0 \\ 0 & 0 \end{bmatrix}.
\end{align*}

If $q \in \{ 0, 1 \}$ then $\mathcal{C}$ is unitarily correctable.
Otherwise, $\mathcal{C}$ is rank-$2$ correctable. The image
algebra under the action of
$\mathcal{E}\circ\mathcal{P}_{\mathcal{C}}$ is given by the
operators of the form
\begin{align}\label{eq:alg01}
 \begin{bmatrix}U\rho U^{\dagger} & 0 \\ 0 & W\rho W^{\dagger}\end{bmatrix},
\end{align}

\noindent where $\rho \in M_{2}$. Moreover,
\begin{align*}
  \mathcal{E}^{\dagger}\Big( \begin{bmatrix}U\rho U^{\dagger} & 0 \\ 0 & W\rho W^{\dagger}\end{bmatrix} \Big) = \begin{bmatrix}\rho & 0 \\ 0 & q^{2}V^{\dagger}W\rho W^{\dagger}V + (1-q^2)\rho\end{bmatrix},
\end{align*}

\noindent from which it follows that $\mathcal{E}^{\dagger}$ is a
$*$-homomorphism when restricted to this algebra if and only if $q
\in \{ 0, 1 \}$ (in which case $\mathcal{C}$ is unitarily
correctable) or $W = V$. It is not difficult to verify, however,
that $W = V$ is exactly the condition under which
$\mathcal{L}(\mathcal{H})$ becomes a unitarily correctable
subsystem when the space is decomposed as $M_2 \otimes
\mathcal{L}(\mathcal{H})$. Further, the image of the algebra
$1_{\mathcal{A}}\otimes \mathcal{L}(\mathcal{H})$ under
$\mathcal{E}$ is exactly the algebra of operators of the form in
Eq.(~\eqref{eq:alg01}). }
\end{exam}

It is also worth noting that if $\mathcal{E}$ is not unital, then
Theorem~\ref{thm:multDom2} does not hold, even just when
considering $MD(\mathcal{E}^{\dagger})$ and
$\mathcal{E}(UCC(\mathcal{E}))$. This can be seen explicitly by
the following example, which gives a non-unital channel
$\mathcal{E}$ with a noiseless subspace that is not captured under
the image of $\mathcal{E}$ by the multiplicative domain of
$\mathcal{E}^{\dagger}$. Nevertheless, it will be seen in
Theorem~\ref{thm:multDom3} that the multiplicative domain can help
us find a subclass of unitarily correctable codes for non-unital
quantum channels.

\begin{exam}\label{exam:multDom1}
{\rm Let $q \in [0, \frac{1}{2}]$ and define a quantum channel
$\mathcal{E}$ on a $4$-dimensional Hilbert space $\mathcal{H}$ by
the following $3$ Kraus operators in the standard basis:
\begin{align*}
E_{1} = \begin{bmatrix} \alpha & 0 & 0 & 0 \\ 0 & 1 & 0 & 0 \\ 0 & 0 & 1 & 0 \\ 0 & 0 & 0 & \alpha \end{bmatrix} \ \ \ \ E_{2} = \beta \begin{bmatrix} 1 & 0 & 0 & 0 \\ 0 & 0 & 0 & 1 \\ 1 & 0 & 0 & 0 \\ 0 & 0 & 0 & 1 \end{bmatrix} \ \ \ \ E_{3} = \beta \begin{bmatrix} 1 & 0 & 0 & 0 \\ 0 & 0 & 0 & -1 \\ -1 & 0 & 0 & 0 \\ 0 & 0 & 0 & 1 \end{bmatrix},
\end{align*}

\noindent where $\alpha = \sqrt{1 - 2q}$ and $\beta = \sqrt{q/2}$. It is straightforward to verify that $\mathcal{E}$ is a nonunital quantum channel. It is similarly not difficult to verify that a decoherence-free subspace of dimension $2$ for $\mathcal{E}$ is projected onto by the projection
\begin{align*}
  P_{\mathcal{C}} = \begin{bmatrix} 0 & 0 & 0 & 0 \\ 0 & 1 & 0 & 0 \\ 0 & 0 & 1 & 0 \\ 0 & 0 & 0 & 0\end{bmatrix}.
\end{align*}

The image algebra under the action  of
$\mathcal{E}\circ\mathcal{P}_{\mathcal{C}}$ is then simply
$\mathcal{L}_1(P_{\mathcal{C}}\mathcal{H})$. Observe that
\begin{align*}
  \mathcal{E}^{\dagger}\Big( \begin{bmatrix}0 & 0 & 0 & 0 \\ 0 & r & s & 0 \\ 0 & t & u & 0 \\ 0 & 0 & 0 & 0\end{bmatrix} \Big) = \begin{bmatrix}0 & 0 & 0 & 0 \\ 0 & r & s & 0 \\ 0 & t & u & 0 \\ 0 & 0 & 0 & 0\end{bmatrix} + q\begin{bmatrix}u & 0 & 0 & t \\ 0 & 0 & 0 & 0 \\ 0 & 0 & 0 & 0 \\ s & 0 & 0 & r\end{bmatrix},
\end{align*}

\noindent from which it follows that $\mathcal{E}^{\dagger}$ is a
$*$-homomorphism when restricted to this algebra if and only if $q
= 0$ (in which case $\mathcal{E}$ is unital) or $q = 1$ (in which
case $\mathcal{E}$ is not trace-preserving). }
\end{exam}

For an arbitrary non-unital channel $\mathcal{E}$, it is not at
all clear how one could go about computing its UCCs. For instance,
there does not appear to be an analogue of the algebra
$UCC(\mathcal{E})$ in the general non-unital case. However, the
following theorem shows how the previous results on the
multiplicative domain can be extended to the non-unital case, and
hence that it yields a subclass of UCCs that can be directly
computed. On terminology, when we say the ``codes encoded in an
algebra'', we mean the subsystem (and subspace) codes determined
by the structure of the algebra as in Eq.~(\ref{opalgform}).

\begin{thm}\label{thm:multDom3}
  Let $\mathcal{E}$ be a quantum operation. Then the quantum codes encoded in
  $MD(\mathcal{E})$ are UCC for $\mathcal{E}$.
\end{thm}
\begin{proof}
Proceeding similarly to the proof of Theorem~\ref{thm:multDom2},
note that if $a \in MD(\mathcal{E})$ then
$\Tr(\mathcal{E}(a)\mathcal{E}(b)) = \Tr(\mathcal{E}(ab))$ for all
$b \in \mathcal{L}_1(\mathcal{H})$. Thus
$\Tr(\mathcal{E}^{\dagger}\circ\mathcal{E}(a)b) =
\Tr(\mathcal{E}^{\dagger}(I)ab) = \Tr(ab)$ for all $b \in
\mathcal{L}_1(\mathcal{H})$ and so it follows that
$\mathcal{E}^{\dagger}\circ\mathcal{E}(a) = a$ for all $a \in
MD(\mathcal{E})$. The remainder of this proof shows that this
implies that $a$ is contained in a unitarily correctable subsystem
of $\mathcal{E}$.

Assume without loss of generality that $a$ is of the form
$I_{\mathcal{A}} \otimes \rho_{\mathcal{B}}$. Then we have that
$\mathcal{E}^{\dagger}\circ\mathcal{E}(I_{\mathcal{A}}\otimes
\rho_{\mathcal{B}}) = I_{\mathcal{A}}\otimes\rho_{\mathcal{B}}$
for all $\rho_{\mathcal{B}}$. This implies from the positivity and
linearity of $\mathcal{E}^{\dagger}\circ\mathcal{E}$ that for any
$\sigma_\mathcal{A}$ there is a  $\tau_\mathcal{A}$ such that
$\mathcal{E}^{\dagger}\circ\mathcal{E}(\sigma_{\mathcal{A}}\otimes
\rho_{\mathcal{B}}) = \tau_{\mathcal{A}}\otimes\rho_{\mathcal{B}}$
for all $\rho_{\mathcal{B}}$. Thus, multiplying on the left by
$\mathcal{P}_\mathcal{C}$ gives us
$\mathcal{P}_\mathcal{C}\circ\mathcal{E}^{\dagger}\circ\mathcal{E}\circ\mathcal{P}_\mathcal{C}
= (\mathcal{F}_\mathcal{A}\otimes{\rm
id}_\mathcal{B})\circ\mathcal{P}_\mathcal{C}$ for some channel
$\mathcal{F}_\mathcal{A}$, and hence $\mathcal{B}$ is correctable
for $\mathcal{E}$.

It then follows from Lemma~\ref{lem:coderep02} that
$\sum_{i}{(D_{i}^{4} \otimes \rho_{\mathcal{B}})} =
I_{\mathcal{A}} \otimes \rho_{\mathcal{B}}$. Hence
$\sum_{i}{D_{i}^{4}} = I_{\mathcal{A}}$, and in particular $d_{ij}
\leq 1$ for all $i,j$ where $d_{ij}$ is the $j^{th}$ diagonal
entry of $D_{i}$ in a diagonal matrix representation (recall the
$D_i$ are mutually commuting and hence simultaneously
diagonalizable). Also, it comes out of the proof of that lemma
that $\sum_{i}D_{i}^{2} = I_{\mathcal{A}}$. It then follows that
exactly $\dim(\mathcal{A})$ of the $d_{ij}$ equal $1$, and the
rest equal $0$. Now apply the procedure used to prove
Lemma~\ref{lem:coderep02}, while being sure to pick the unitary
$U$ so that it permutes all of the diagonal entries of $D =
(D_{ij})$ to the top-left block. Doing this will ensure that the
channel $\mathcal{G}$ has only a single Kraus operator, and thus
$B$ must be unitarily correctable for $\mathcal{E}$.
\end{proof}

Note that one thing that comes out of the proof of this result is
that the implication (2) $\Rightarrow$ (1) of
Theorem~\ref{thm:kribSpek1} holds for non-unital channels as long
as $\mathcal{E}^{\dagger}\circ\mathcal{E}(P_{\mathcal{C}}) =
P_{\mathcal{C}}$. In particular, that implication always holds for
noiseless and unitarily correctable \emph{subspaces}.

\begin{exam}
{\rm We give a simple example of a channel with a non-trivial
multiplicative domain that does not capture all UCCs. Let
$\mathcal{E}$ be the channel defined on $6 \times 6$ matrices,
broken up into nine $2\times 2$ blocks, as follows:
\[
\mathcal{E} \left[ \begin{matrix} A_{11} & A_{12} & A_{13} \\
A_{21} & A_{22} & A_{23} \\
A_{31} & A_{32} & A_{33} \\
\end{matrix} \right] = \left[ \begin{matrix} 0 & 0 & 0 \\ 0 &
A_{11} + A_{22} & 0 \\ 0 & 0 & A_{33}
\end{matrix} \right] .
\]
Clearly each of the three block entries $(i,i)$, $i=1,2,3$, define
single qubit unitarily correctable codes, but only the third is
encoded in the multiplicative domain. In fact, in this case the
$2\times 2$ block determined by the $(3,3)$ entry is precisely the
multiplicative domain for $\mathcal{E}$.
 }
\end{exam}

\begin{remark}
{\rm This example is very much in the spirit of the spontaneous
emission or amplitude dampening channels \cite{NC00}, which are
the standard physical examples of  non-unital quantum channels. It
would be interesting to know if the non-unital behaviour of
arbitrary channels could somehow be characterized by such
channels, and what role, if any, the multiplicative domain might
have in the description. We plan to undertake this investigation
elsewhere. }
\end{remark}

\subsection{Computing The Multiplicative Domain}

While it is not known how to compute UCC for an arbitrary channel,
the multiplicative domain codes can be computed with available
software. In order to compute the multiplicative domain of a
linear map $\phi : M_{n} \mapsto M_{k}$, note that it suffices to
solve the following system of $2k^{2}n^{2}$ linear equations in
$n^2$ unknowns:
\begin{align*}
  \phi(E_{l,m}(\sigma_{i,j})) = \phi(E_{l,m})\phi((\sigma_{i,j})) \\ \text { and } \quad \quad \quad \quad \quad \quad \\ \phi((\sigma_{i,j})E_{l,m}) = \phi((\sigma_{i,j}))\phi(E_{l,m}),
\end{align*}

\noindent for all $1\leq l,m \leq n$, where $\big\{ E_{l,m}
\big\}$ is the family of standard matrix units associated with a
fixed basis. If we let $\phi = \big\{ A_{p} \big\}$, where $A_{p}
= (a_{ijp})$ (where $i$ indexes the rows of $A_p$ and $j$ indexes
the columns of $A_p$), then the above matrix equations can be
written out more explicitly as the following system of linear
equations
\begin{align*}
  \sum_{b,e}{a_{ywe}\overline{a_{zbe}}\sigma_{xb}} = \sum_{b,c,d,e,f}{a_{ywe}\overline{a_{dxe}}a_{dcf}\overline{a_{zbf}}\sigma_{cb}} \quad \forall \, 1 \leq w,x \leq n, 1 \leq y,z \leq k
\end{align*}
and
\begin{align*}
  \sum_{b,e}{a_{ybe}\overline{a_{zxe}}\sigma_{bw}} = \sum_{b,c,d,e,f}{a_{yce}\overline{a_{dbe}}a_{dwf}\overline{a_{zxf}}\sigma_{cb}} \quad \forall \, 1 \leq w,x \leq n, 1 \leq y,z \leq k.
\end{align*}

This is simply a system of linear equations and thus can be solved
by computer software such as MATLAB. For large scale quantum
systems, however, it is clear that more refined approaches would
be  required to compute these (as well as any other) codes. We
leave such scalability issues for investigation elsewhere.

\begin{exam}
{\rm  This example illustrates how the above linear system of
equations can be used to compute the multiplicative domain of an
arbitrary map, and find unitarily correctable codes from it.
Consider again the channel from Example~\ref{exam:subspace}, but
choose $U = V = I$. That is, consider the $2$-qubit channel
$\mathcal{E}$ defined by the four Kraus operators
  \begin{align*}
\alpha\begin{bmatrix} I & I \\ 0 & 0 \end{bmatrix}, \quad
\alpha\begin{bmatrix} I & -I \\ 0 & 0 \end{bmatrix} , \quad
\beta\begin{bmatrix} I & I \\ I & I
\end{bmatrix}, \quad \beta\begin{bmatrix} -I & I \\ I & -I
\end{bmatrix},
\end{align*}

\noindent where $\alpha = \frac{\sqrt{q}}{\sqrt{2}}$, $\beta =
\frac{\sqrt{1 - q}}{2}$, and $q \in [0,1]$. Then if we write
$\sigma = \begin{bmatrix}A & B \\ C & D\end{bmatrix}$, where
$A,B,C,D\in M_2$ are $2 \times 2$ matrices, then the linear
equations that need to be solved reduce to
\begin{eqnarray*}
  (1-q)A  = (1+q)D \quad\quad (1-q)B  = (1+q)C \\
  (1+q)A  = (1-q)D \quad\quad (1+q)B  = (1-q)C.
\end{eqnarray*}
We will consider the solutions of these linear equations in three cases. \\

\noindent{\bf Case 1:} $q = 0$. In this case the solutions are $A
= D$ and $B = C$, so the multiplicative domain of $\mathcal{E}$
consists of exactly the matrices of the form $\begin{bmatrix}A & B
\\ B & A\end{bmatrix}$. Because this channel is unital when $q =
0$, it follows by Theorem~\ref{thm:multDom2} that the algebra of
unitarily correctable codes is exactly the same,
\begin{align}\label{eqn:alg01}
  UCC(\mathcal{E}) = \Big\{ \begin{bmatrix}A & B\\ B & A\end{bmatrix} : A,B\in M_2\Big\}.
\end{align}
Indeed, it is not difficult to verify that this algebra encodes,
in the sense discussed above, a pair of decoherence-free subspaces
for $\mathcal{E}$.\\

\noindent{\bf Case 2:} $0 < q < 1$. The solutions here are $A = B
= C = D = 0$, so the multiplicative domain contains only the zero
matrix and thus does not capture any correctable codes. It appears
these channels also do not have unitarily correctable codes, though they
do have rank-2 correctable codes as described in Example~\ref{exam:subspace}. \\

\noindent{\bf Case 3:} $q = 1$. The solutions here are $A = B = C
= D = 0$, so the multiplicative domain contains only the zero
matrix and thus does not capture any correctable codes. However,
it is easily verified that the two subspaces defined by the ranges
of the following two algebras are unitarily correctable:
\begin{align*}
  \Big\{ \begin{bmatrix}A & -A\\ -A & A\end{bmatrix} : A\in M_2\Big\}\quad\text{ and }\quad \Big\{ \begin{bmatrix}A & A\\ A & A\end{bmatrix} : A\in M_2\Big\},
\end{align*}

\noindent where the unitary correction operations are
$\frac{1}{\sqrt{2}}\begin{bmatrix}I & I \\ -I & I\end{bmatrix}$
and $\frac{1}{\sqrt{2}}\begin{bmatrix}I & -I \\ I &
I\end{bmatrix}$, respectively. The fact that the multiplicative
domain does not capture all unitarily correctable codes highlights
the fact that the converse of Theorem~\ref{thm:multDom3} does not
hold in general for non-unital quantum channels. Also, the
smallest algebra containing these two subspaces is exactly the
algebra described by Eq.~(\ref{eqn:alg01}). However,
\begin{align*}
 \mathcal{E}\Big(\begin{bmatrix}A & B \\ B & A\end{bmatrix}\Big) = \begin{bmatrix}2A & 0 \\ 0 & 0\end{bmatrix},
\end{align*}

\noindent so clearly that algebra is not unitarily correctable as
there is no way to recover the ``$B$'' blocks. This highlights the
fact that in general there is no way to define the UCC algebra of
a non-unital quantum channel. }
\end{exam}

\vspace{0.1in}

\noindent{\bf Acknowledgements.} M.-D.C. was supported by NSERC
Discovery Grant. N.J. was supported by an NSERC Canada Graduate
Scholarship and the University of Guelph Brock Scholarship. D.W.K.
was supported by NSERC Discovery Grant and Discovery Accelerator
Supplement, an Ontario Early Researcher Award, and CIF, OIT.



\begin{thebibliography}{99}

\bibitem{NC00} M.~A. Nielsen, I.~L. Chuang,
\textit{Quantum computation and quantum information}, Cambridge
University Press, 2000.

\bibitem{Got02} D. Gottesman, \emph{An introduction to quantum error correction},
in Quantum Computation: A Grand Mathematical Challenge for the
Twenty-First Century and the Millennium, ed. S. J. Lomonaco, Jr.,
pp. 221-235 (American Mathematical Society, Providence, Rhode
Island, 2002).

\bibitem{KLABVZ02} E. Knill, R. Laflamme, A. Ashikhmin, H.N. Barnum, L. Viola,
W.H. Zurek, \emph{Introduction to quantum error correction}, Los
Alamos Science, November 27, 2002.

\bibitem{Got96}
D. Gottesman, Phys. Rev. A {\bf 54},  1862  (1996).

\bibitem{KLP05} D. Kribs, R. Laflamme, D. Poulin, Phys. Rev. Lett. {\bf 94},
180501 (2005).

\bibitem{Pou05} D. Poulin, Phys. Rev. Lett. {\bf 95}, 230504 (2005).

\bibitem{SL05} A. Shabani, D.A. Lidar, Phys. Rev. A, {\bf 72}
042303 (2005).

\bibitem{Bac05} D. Bacon, Phys. Rev. A, {\bf 73} 012340 (2006).

\bibitem{KS06} D.~W. Kribs, R.~W. Spekkens, Phys. Rev. A {\bf 74},
042329 (2006).

\bibitem{CK06} M.~D. Choi, D.~W. Kribs, Phys. Rev. Lett., {\bf 96} 050501 (2006).

\bibitem{Kni06} E. Knill, Phys. Rev. A {\bf 74}, 042301 (2006).

\bibitem{KlaSar06} A.~Klappenecker, P.~K. Sarvepalli, arXiv.org/quant-ph/0604161 (2006).

\bibitem{AlyKla07} S.~A. Aly, A. Klappenecker, arXiv:0712.4321 (2007).

\bibitem{ESMRLBCL07} J. Emerson, M. Silva, O. Moussa, C. Ryan, M.
Laforest, J. Baugh, D.~G. Cory, R. Laflamme, Science {\bf 317},
1893 (2007).

\bibitem{BKK07a} C. Beny, A. Kempf, D.~W. Kribs, Phys. Rev. Lett., {\bf 98} 100502 (2007).

\bibitem{BNPV08} R. Blume-Kohout, H.K. Ng, D. Poulin, L. Viola, Phys. Rev. Lett., {\bf 100} 030501 (2008).

\bibitem{SMKE08} M. Silva, E. Magesan, D.~W. Kribs, J. Emerson,
Phys. Rev. A, {\bf 78} 012347 (2008).

\bibitem{Cho74} M.-D. Choi, Illinois J. Math., {\bf 18} (1974), 565-574.

\bibitem{Paulsentext} V. I. Paulsen,
{\em Completely Bounded Maps and Operator Algebras,}
Cambridge University Press, Cambridge, 2003.

\bibitem{KLPL06} D.~W. Kribs, R. Laflamme, D. Poulin, M. Lesosky,
Quantum Inf. \& Comp. {\bf 6} (2006), 383-399.

\bibitem{zanardi97}
P. Zanardi, M. Rasetti, Phys. Rev. Lett. {\bf 79},  3306 (1997).

\bibitem{palma96}
G. Palma, K.-A. Suominen,  A. Ekert, Proc. Royal Soc. A {\bf 452},
567  (1996).

\bibitem{duan97}
L.-M. Duan  G.-C. Guo, Phys. Rev. Lett. {\bf 79},  1953 (1997).

\bibitem{lidar98}
D.~A. Lidar, I.~L. Chuang, K.~B. Whaley, Phys. Rev. Lett. {\bf
81},  2594  (1998).

\bibitem{knill00}
E. Knill, R. Laflamme, L. Viola, Phys. Rev. Lett. {\bf 84}, 2525
(2000).

\bibitem{zanardi01a}
P. Zanardi, Phys. Rev. A {\bf 63},  12301  (2001).

\bibitem{kempe01}
J. Kempe, D. Bacon, D.~A. Lidar, K.~B. Whaley, Phys. Rev. A {\bf
63},
  42307  (2001).

\bibitem{shor95}
P.~W. Shor.
Phys. Rev. A \textbf{52}, R2493 (1995).

\bibitem{steane96}
A.~M. Steane.
Phys. Rev. Lett. \textbf{77}, 793 (1996).

\bibitem{bennett96}
C.~H. Bennett, D.~P. DiVincenzo, J.~A. Smolin, W.~K. Wootters.
Phys. Rev. A \textbf{54}, 3824 (1996).

\bibitem{KL97a} E.~Knill, R.~Laflamme, Phys. Rev. A {\bf 55}, 900 (1997).

\bibitem{NP05} M.~A. Nielsen, D. Poulin, Phys. Rev. A \textbf{75}, 64304
(2007).

\bibitem{Dav96}
K.~R. Davidson, {\it $C^*$-algebras by example}, Fields Institute
Monographs, 6. American Mathematical Society, Providence, RI,
1996.

\bibitem{BKP08} C. Beny, D.~W. Kribs, A. Pasieka, Int. J. Quantum
Inf., {\bf 6} (2008), 597-603.

\bibitem{BKK08} C. Beny, A. Kempf, D.~W. Kribs, preprint, 2008.






\end{thebibliography}
\end{document}